\documentclass[aps,onecolumn,showpacs,amsfonts,eqsecnum,float,preprint]{revtex4} 
\usepackage{epsfig,amsmath}
\usepackage{bm}

\begin{document}

\title{Unified derivation of Bohmian methods and the incorporation of interference effects}

\author{ Yair Goldfarb, Jeremy Schiff and David J. Tannor} \affiliation{Dept. of
Chemical Physics, The Weizmann Institute of Science, Rehovot, 76100
Israel\\ \today}

\begin{abstract}

\noindent We present a unified derivation of Bohmian methods that
serves as a common starting point for the derivative propagation
method (DPM), Bohmian mechanics with complex action (BOMCA) and the
zero-velocity complex action method (ZEVCA). The unified derivation
begins with the ansatz $\psi=e^{\frac{iS}{\hbar}}$ where the action,
$S$, is taken to be complex and the quantum force is obtained by
writing a hierarchy of equations of motion for the phase partial
derivatives. We demonstrate how different choices of the trajectory
velocity field yield different formulations such as DPM, BOMCA and
ZEVCA. The new derivation is used for two purposes. First, it serves
as a common basis for comparing the role of the quantum force in the
DPM and BOMCA formulations. Second, we use the new derivation to
show that superposing the contributions of \textit{real},
\textit{crossing} trajectories yields a nodal pattern essentially
identical to that of the exact quantum wavefunction. The latter
result suggests a promising new approach to deal with the
challenging problem of nodes in Bohmian mechanics.

\end{abstract}

\maketitle



\section{Introduction}

\noindent The contrast between the inherent non-locality of quantum
mechanics and the locality of classical mechanics has driven a
decades long quest for a trajectory-based formulation of quantum
theory that is exact. In the 1950's, David Bohm, building on earlier
work by Madelung\cite{madelung} and de Broglie\cite{deBroglie},
developed an exact formulation of quantum mechanics in which
trajectories evolve in the presence of the usual Newtonian force
plus an additional quantum force\cite{bohm,bohmb}. Bohm's
formulation was originally developed as an \textit{interpretational}
tool, to recover a notion of causality in quantum mechanics. In 1999
Lopreore and Wyatt\cite{courtney} demonstrated that the Bohmian
formulation can also be used a \textit{numerical} tool to do quantum
calculations. This innovation has coincided well with the ongoing
interest of the chemical physics community in finding effective
numerical tools for performing multidimensional quantum
calculations. The apparently \textit{local} dynamics of the Bohmian
trajectories suggest the possibility of computational advantages
compared to fixed grid methods or direct-product basis set methods
that scale exponentially with dimensionality. The Lopreore and Wyatt
paper has motivated the development of a variety of new numerical
approaches for implementing Bohmian
mechanics\cite{jian,erik,sophya,burgha,ginden,poirier,kend}.
Reference \cite{wyattb} gives an excellent account of the progress
in Bohmian related formulations in recent years.

The Bohmian formulation has two main limitations that currently
limit its usefulness as a numerical tool. First, note that the
non-locality of quantum mechanics does not disappear in the Bohmian
formulation---it manifests itself in the quantum force term, which
needs to be calculated to propagate the quantum trajectories. The
quantum force is very unstable numerically when the wavefunction is
oscillatory. This is related to the second limitation: the quantum
force diverges at nodes of the wavefunction resulting in the break
down of the Bohmian formulation in the vicinity of nodes. Indeed,
most of the new contributions to Bohmian methodology have been aimed
at overcoming these two limitations of the Bohmian formulation.

Among the approaches that have been developed to deal with the
calculation of the quantum force are the derivative propagation
method (DPM)\cite{corey} and Bohmian mechanics with complex action
(BOMCA)\cite{goldfarb,goldfarb3}. The two approaches use a similar
procedure writing a hierarchy of equations of motion for partial
derivatives of the phase. The difference between the methods lies in
the use of a \textit{complex} action in BOMCA as opposed to a real
amplitude and phase in DPM. This difference has far-reaching
consequences that we explore in this paper.

In the first part of this publication we present a unified and
compact derivation of Bohmian methods that serves as a common
starting point for the DPM and BOMCA methods, and the Zero-Velocity
Complex Action (ZEVCA) method. The derivation is similar to the
derivation of BOMCA but it leaves the choice of the trajectory
velocity field undetermined. We demonstrate how different choices of
the velocity field yield different formulations, such as DPM, BOMCA
and others. The unified derivation allows for ready comparison of
DPM and BOMCA, and the role that the quantum force plays in both
formulations. In the second part of this paper we use the new
derivation to show that superposing the contributions of
\textit{real}, \textit{crossing} trajectories yields an
\textit{interference} pattern. This contrasts with a cardinal
principle of the exact Bohmian formulation, that trajectories are
not allowed to cross in configuration space. However, since we are
dealing with an \textit{approximation} to Bohmian mechanics, the
no-crossing role does not apply. Superposing the contribution from
real crossing trajectories represents a promising new avenue for
dealing with the challenging problem of nodes in Bohmian mechanics.

This paper is organized as follows. In section \ref{general} we
present the unified derivation. Section \ref{specific} is dedicated
to several special cases of the derivation: ZEVCA (\ref{zevca}),
BOMCA (\ref{bomca}) and DPM (\ref{dpm}). In Section \ref{inter} we
show how a variation of the derivation leads to an accurate
description of interference and nodal patterns. Section
\ref{summary} is a summary and concluding remarks.


\section{Unified derivation of Bohmian methods}
\label{general}

\noindent We start by inserting the ansatz\cite{pauli,kurt,david}
\begin{equation}
\label{trial2} \psi(x,t)=\exp\left[\frac{i}{\hbar}S(x,t)\right],
\end{equation}
into the TDSE
\begin{equation}
i\hbar\psi_{t}=-\frac{\hbar^{2}}{2m}\psi_{xx}+V(x,t)\psi,
\end{equation}
where $S(x,t)$ is a complex function, $\hbar$ is Planck's constant
divided by $2\pi$, $m$ is the mass of the particle and $V(x,t)$ is
the potential energy function. The subscripts denote partial
derivatives. The result is a quantum complex Hamilton-Jacobi (HJ)
equation\cite{pauli,kurt,david}
\begin{equation}
\label{st1}
 S_{t}+\frac{1}{2m}S^{2}_{x}+V=\frac{i\hbar}{2m}S_{xx},
\end{equation}
where we recognize on the LHS a \textit{complex} version of the
classical HJ equation. On the RHS is an additional non-classical
term that we refer to as the ``complex quantum potential". This term
is different from the quantum potential in the conventional Bohmian
formulation. The \textit{time-independent} version of ansatz
(\ref{trial2}) is familiar as the starting point of the WKB
approximation\cite{wentzel,kramers,brillouin}. Equation (\ref{st1})
was derived by Pauli\cite{pauli} as a first step in the WKB
derivation. More recently, eq.(\ref{st1}) was rediscovered by
Leacock and Padgett\cite{leacock}, who also reverted to a
time-independent formulation to calculate eigenvalues. The equation
has also been used by several other authors\cite{john,yang,san} as
an analytical tool, but not as a constructive method to solve the
TDSE with trajectories.

The classical HJ equation can be conveniently solved by integrating
along trajectories that satisfy the classical equations of motion.
From a mathematical point of view, these trajectories are the
characteristics of the classical HJ equation. Here we use an
analogous approach to solve the quantum complex HJ equation, by
integrating along some family of trajectories.  We define a family
of trajectories by \textit{choosing} a velocity field $v(x,t)$,
which can be done in an infinite number of ways. (As opposed to the
classical case, the velocity field is not necessarily predetermined
by the PDE we are trying to solve.) Solutions of
\begin{equation}
\label{dxdt} \frac{dx}{dt}=v(x,t)
\end{equation}
determine trajectories, parametrized by their initial position
$x(0)$. The change of the phase \textit{along} the trajectory is
given by operating on $S(x,t)$ with the Lagrangian time derivative
\begin{equation}
\frac{d}{dt}\equiv\frac{\partial}{\partial
t}+\frac{dx}{dt}\frac{\partial}{\partial x}
=\frac{\partial}{\partial t}+v\frac{\partial}{\partial x},
\end{equation}
with the result
\begin{equation}
\label{dsdt}
\frac{dS}{dt}=S_{t}+vS_{x}=\frac{i\hbar}{2m}S_{xx}-\frac{1}{2m}S^{2}_{x}-V+vS_{x},
\end{equation}
where we have used eq.(\ref{st1}). The integration of
eq.(\ref{dsdt}) requires $S_{x}[x(t),t]$ and $S_{xx}[x(t),t]$, i.e.
the values of the first two spatial derivatives of $S$ along the
trajectory. Fortunately, it is possible to write equations
describing the evolution of the spatial derivatives of $S$. Writing
\begin{equation}
S_{n}[x(t),t]\equiv \left.\frac{\partial^{n} S}{\partial
x^{n}}\right|_{[x(t),t]},
\end{equation}
and taking the $n^{th}$ spatial derivative of eq.(\ref{st1}), we
have
\begin{equation}
\label{st2}
 (S_{t})_{n}+\frac{1}{2m}(S^{2}_{1})_{n}+V_{n}=\frac{i\hbar}{2m}S_{n+2}\
 .
\end{equation}
Inserting the result in the definition of the Lagrangian time
derivative of $S_{n}$ gives
\begin{equation}
\label{dsndt1} \frac{dS_{n}}{dt}=(S_{t})_{n}+vS_{n+1}=
-\frac{1}{2m}(S^{2}_{1})_{n}-V_{n}+vS_{n+1}+\frac{i\hbar}{2m}S_{n+2}\
.
\end{equation}
From this procedure we obtain an infinite set of coupled equations
describing the evolution of $S$ and its spatial derivatives along a
trajectory, namely eqs.({\ref{dsndt1}) for $n=0,1,2,\ldots$. We note
that evaluation of $S_{n}$ requires knowledge of $S_{n+2}$ and
$S_{n+1}$ (the latter through the term
$(S_{1}^{2})_{n}=\sum_{j=0}^{n}\binom{n}{j}S_{j+1}S_{n-j+1}$). The
fact that evaluation of $S_{0}=S$ requires knowledge of $S_2=S_{xx}$
expresses the \textit{nonlocality} of the Schr\"odinger equation.
However from a formal point of view, integration of the infinite
hierarchy of eqs.(\ref{dsndt1}) along the trajectories defined by
eq.(\ref{dxdt}) can be regarded as a local method of solution of
eq.(\ref{st1})in the sense that information propagates along
individual trajectories independently. This does not contradict the
nonlocality of the Schr\"odinger equation, since not only is the
value of $S$ being propagated down a trajectory, but also all its
spatial derivatives $S_1,S_2,\ldots$ are being propagated.

A numerical approximation for solving eq.(\ref{st1}) can be obtained
by truncating the set (\ref{dsndt1}) at some $n=N$, by setting
$S_{N+1}=S_{N+2}=0$. We summarize the equations of motion of the
approximation:
\begin{eqnarray}
\label{dxdt2} \frac{dx}{dt}&=&v[x(t),t], \\ \label{st3}
\frac{dS_{n}}{dt}&=&-\frac{1}{2m}(S^{2}_{1})_{n}-V_{n}+vS_{n+1}+\frac{i\hbar}{2m}S_{n+2};
\ \ n=0,1,...,N,
\\ \label{sn1}
S_{N+1}&=&S_{N+2}=0,
\end{eqnarray}
where we emphasize that the solution is given along an individual
trajectory $x(t)$ determined by the velocity field $v(x,t)$. The
initial conditions of eqs.(\ref{st3}) are given by
\begin{equation}
\label{initial}
S_{n}[x(0),0]=-i\hbar\left.\frac{\partial^{n}\ln[\psi(x,0)]}{\partial
x^{n}}\right|_{x(0)}.
\end{equation}
where we have used the relation $S(x,0)=-i\hbar\ln[\psi(x,0)]$ from
ansatz (\ref{trial2}). The wavefunction at time $t_{f}$ at final
position $x(t_{f})$ is given by
\begin{equation}
\label{finalwf}
\psi[x(t_{f}),t_{f}]=\exp\left\{\frac{i}{\hbar}S_{0}[x(t_{f}),t_{f}]\right\}.
\end{equation}
Equations (\ref{dxdt2})-(\ref{sn1}) provide a common starting point
for deriving several quantum trajectory methods such as DPM, BOMCA
and ZEVCA as well as adaptive grid techniques. These methods differ
from each other by the specific choice of the velocity field. For
the sake of simplicity we present and compare the different methods
for $N=2$. For this case, eqs.(\ref{dxdt2})-(\ref{sn1}) yield four
equations of motion
\begin{subequations}
\label{N2}
\begin{align}
\frac{dx}{dt}&=v,   \\
\frac{dS_{0}}{dt}&=-\frac{S_{1}^{2}}{2m}-V+vS_{1}+\frac{i\hbar}{2m}S_{2},  \\
\frac{dS_{1}}{dt}&=-\frac{S_{1}}{m}S_{2}-V_{1}+vS_{2}, \\
\frac{dS_{2}}{dt}&=-\frac{S_{2}^{2}}{m}-V_{2}.
\end{align}
\end{subequations}
The $N=2$ case is unique for two reasons. First, it is the lowest
order of truncation for which the equation of motion for the phase
$S_{0}$ (eq.(\ref{st3}) for $n=0$) includes a quantum potential term
$\frac{i\hbar}{2m}S_{2}$. Second, eqs.(\ref{N2}) for BOMCA, ZEVCA
and DPM yield the \textit{exact} solution for an initial Gaussian
wavepacket propagating in a potential with up to quadratic terms.


\section{Specific choices of trajectory velocity fields}
\label{specific}

\subsection{Zero-Velocity Complex Action (ZEVCA)}
\label{zevca}

\noindent The simplest choice of the velocity field is
\begin{equation}
v(x,t)=0\Longrightarrow x(t)=x(0).
\end{equation}
The resulting trajectories are straight lines, hence we refer to the
resulting approximation as the Zero-Velocity Complex Action
method\cite{goldfarb2} (ZEVCA). The ZEVCA formulation can be
regarded as a hybrid between a grid method and a local semiclassical
method. In reference \cite{wyattb} section 7.2, Wyatt considers the
solution of the global hydrodynamic equations of quantum mechanics
on fixed grid points (Eulerian grid) but dismisses its usefulness as
a numerical tool. In reference \cite{goldfarb2} the ZEVCA
formulation is shown to produce useful output from local propagation
at a single grid point. In this paper we focus on the relation
between BOMCA and DPM, hence we will not elaborate further here on
the ZEVCA method. The interested reader is referred to
\cite{goldfarb2}.


\subsection{Bohmian Mechanics with Complex Action (BOMCA)}

\label{bomca}
\noindent
 In the BOMCA method, the velocity of the trajectories is set as
\begin{equation}
\label{v_bomca}
 v[x(t),t]=\frac{S_{1}[x(t),t]}{m}.
\end{equation}
The rationale for this choice is evident if we recall that $S_{1}$
has units of momentum and that $S_{0}$ can be associated with a
quantum action field (eq.(\ref{st1})). Inserting eq.(\ref{v_bomca})
in eqs.(\ref{N2}) yields the $N=2$ equations of motion of BOMCA
\begin{subequations}
\label{bomca_eqs}
\begin{align}
 \frac{dx}{dt}&=\frac{S_{1}}{m}, \\
 \frac{dS_{0}}{dt}&=\frac{S_{1}^{2}}{2m}-V+\frac{i\hbar}{2m}S_{2},
\\
 \frac{dS_{1}}{dt}&=-V_{1}, \\
 \frac{dS_{2}}{dt}&=-V_{2}-\frac{S_{2}^2}{m}.
\end{align}
\end{subequations}
The BOMCA formulation was originally presented in terms of equations
of motion for partial derivatives of the velocity instead of the
phase\cite{goldfarb}. The transformation between the two
formulations is straightforward by using relation (\ref{v_bomca}).

Equations (\ref{bomca_eqs}) have several attractive features. 1)
Equations (\ref{bomca_eqs}a) and (\ref{bomca_eqs}c) are Newton's
second law of motion in disguise, resulting in characteristics that
are \textit{classical} trajectories. In other words, the $N=2$ BOMCA
approximation does not incorporate a quantum force in computing the
trajectories. Note, however, that generally the classical
trajectories take on complex values of position and momentum. This
results from the complex initial momentum that emerges from
eq.(\ref{initial}) for $n=1$. 2) Recognizing the RHS of
eq.(\ref{bomca_eqs}b) as a ``quantum Lagrangian", this equation has
the familiar structure of the equation of motion for the classical
action, with the addition of the quantum potential. The quantum
potential is the only $\hbar$-dependent term, hence this term is
entirely responsible for incorporating the quantum effects in the
$N=2$ BOMCA approximation. Note that for $N=3$, a quantum force term
appears in the \textit{trajectory} equations (specifically
eq.(\ref{bomca_eqs}c)), yielding \textit{complex quantum}
trajectories.

Equations (\ref{bomca_eqs}a-d) appear in the context of other
time-dependent semiclassical methods that use complex classical
trajectories. The first of these is the generalized Gaussian
wavepacket dynamics (GGWPD) of Huber and Heller\cite{huber1,huber2}.
The two main advantages of the BOMCA formulation over GGWPD is that
1) the latter does not provide correction terms to the $N=2$
approximation, and 2) the latter cannot be generalized to a
non-Gaussian initial wavefunction. The second context in which
eqs.(\ref{bomca_eqs}) appear is in the first-order formulation of
the complex trajectory method (CTM) of Boiron and
Lombardi\cite{boiron,goldfarb4}. This formulation is a complex
trajectory version of time-dependent WKB. Although the CTM method is
identical to BOMCA for $N=2$, it differs at higher orders of $N$.

The propagation of complex trajectories requires the analytical
continuation of the initial wavefunction to the complex plane, as
well as a method to reconstruct the wavefunction on the real axis at
the desired final time. Regarding the latter, in
reference\cite{goldfarb} we describe an algorithm to calculate
complex initial positions that end at a final time $t_{f}$ on the
real axis ($\Im[x(t_{f})]=0$). Without going into the details of the
algorithm, suffice it to say that it exploits the local analyticity
of the mapping $x(0)\rightarrow x(t_{f}$). By inserting the phase
$S_{0}$ that corresponds to these trajectories into the original
ansatz (\ref{trial2}), we obtain the wavefunction at a real position
$x(t_{f})$ at time $t_{f}$.


\subsection{Derivative Propagation Method (DPM)}

\label{dpm} \noindent We present here a simplified derivation of the
DPM method that yields an equivalent but more compact set of
equations than has appeared previously in the
literature\cite{wyattb}. Setting the velocity field to be
\begin{equation}
\label{v_dpm}
 v[x(t),t]=\frac{\Re\{S_{1}[x(t),t]\}}{m}
\end{equation}
and inserting eq.(\ref{v_dpm}) into eqs.(\ref{N2}), yields the $N=2$
DPM equations of motion:
\begin{subequations}
\label{dpm_eqs}
\begin{align}
 \frac{dx}{dt}&=\frac{\Re (S_{1})}{m}, \\
 \frac{dS_{0}}{dt}&=\frac{[\Re(S_{1})]^{2}}{2m}-V+\frac{i\hbar}{2m}S_{2}+\frac{[\Im(S_{1})]^{2}}{2m},
\\
 \frac{dS_{1}}{dt}&=-V_{1}-\frac{iS_{2}}{m}\Im(S_{1}), \\
 \frac{dS_{2}}{dt}&=-V_{2}-\frac{S_{2}^2}{m}.
\end{align}
\end{subequations}
The trajectories in the DPM propagate on the real axis, removing the
need to extrapolate the wavefunction to the real axis, as is
required in BOMCA. But the classical structure that was evident in
BOMCA is no longer present: comparing eqs.(\ref{bomca_eqs}) and
(\ref{dpm_eqs}), we see that a quantum force term appears in the
equation of motion for the momentum $S_{1}$ (eq.(\ref{dpm_eqs}c)),
and the equation of motion for the phase $S_{0}$
(eq.(\ref{dpm_eqs}b)) has an ``extra" quantum potential term
$\frac{[\Im(S_{1})]^{2}}{2m}$.

Equations (\ref{dpm_eqs}) are equivalent to those of the DPM as they
appear in the literature\cite{corey,wyattb}, as we now show. In the
DPM, the wavefunction is represented by the ansatz
\begin{equation}
\label{trial3}
 \psi(x,t)=\exp\left[c(x,t)+\frac{i}{\hbar}s(x,t)\right],
\end{equation}
where $s(x,t)$ and $c(x,t)$ are \textit{real} functions (ansatz
(\ref{trial3}) is essentially identical to the conventional Bohmian
ansatz $\psi(x,t)=A(x,t)\exp\left[\frac{i}{\hbar}s(x,t)\right]$ if
one identifies $A(x,t)=\exp[c(x,t)]$). Equating eq.(\ref{trial3})
with eq.(\ref{trial2}) yields
\begin{equation}
\label{ssc} S(x,t)=s(x,t)-i\hbar c(x,t).
\end{equation}
Inserting eq.(\ref{ssc}) into eqs.(\ref{dpm_eqs}) and dividing the
results into their real and imaginary parts yields
\begin{subequations}
\label{dpm_eqs2}
\begin{align}
 \frac{dx}{dt}&=\frac{s_{1}}{m}, \\
 \frac{ds_{0}}{dt}&=\frac{s_{1}^2}{2m}-V+\frac{\hbar^{2}}{2m}(c_{2}+c_{1}^2),
 \\
 \frac{ds_{1}}{dt}&=-V_{1}+\frac{\hbar^{2}}{m}c_{1}c_{2}, \\
 \frac{ds_{2}}{dt}&=-\frac{s_{2}^{2}}{m}+\frac{\hbar^{2}}{m}c_{2}^{2}-V_{2},
 \\
 \frac{dc_{0}}{dt}&=-\frac{s_{2}}{2m}, \\
 \frac{dc_{1}}{dt}&=-\frac{c_{1}s_{2}}{m}, \\
 \frac{dc_{2}}{dt}&=-\frac{2c_{2}s_{2}}{m}.
\end{align}
\end{subequations}
These equations are equivalent to eqs.(\ref{dpm_eqs}a-d), and are
readily seen to be equal to eqs.(10.10) of reference \cite{wyattb}
by taking $N=2$ and identifying $\dot{x}(t)=\frac{s_{1}}{m}$.

Note that the conventional DPM equations of motion also have a
quantum correction to the trajectories at truncating order $N=2$.
This can be seen in the equation for $s_{1}$ (eq.(\ref{dpm_eqs2}c)),
as well as in the equations for $c_{0},c_{1}$ and $c_{2}$
(eqs.(\ref{dpm_eqs2}e-g)) which have no classical counterparts. Even
though the quantum force term in eq.(\ref{dpm_eqs2}c) is
proportional to $\hbar^{2}$, a simple example demonstrates that this
term is not small in any way compared with the classical force.
Consider the harmonic oscillator potential
$V(x)=\frac{1}{2}m\omega^{2}x^{2}$, where $\omega$ is the angular
frequency. The time-evolution of the ground state is
\begin{equation}
\label{sol}
\psi(x,t)=\exp\left[-\alpha_{0}x^2-\frac{i\omega}{2}t+\frac{i\gamma_{0}}{\hbar}\right],
\end{equation}
where $\alpha_{0}=\frac{m\omega}{2\hbar}$ and
$\gamma_{0}=-\frac{i\hbar}{4}\ln(\frac{2\alpha_{0}}{\pi})$. We
recall that eqs.(\ref{dpm_eqs2}) (and eqs.(\ref{bomca_eqs})) yield
the exact solution for an initial Gaussian wavepacket in a harmonic
potential. Comparing eq.(\ref{trial3}) and eq.(\ref{sol}) shows that
$s$ is position-\textit{independent}, hence, $s_{1}=0$. As a
results, according to eq.(\ref{dpm_eqs2}a) the eigen-trajectories
are straight lines, a result familiar from the conventional Bohmian
formulation. Since $\frac{ds_{1}}{dt}=0$, we conclude from
eq.(\ref{dpm_eqs2}c) that the quantum force is equal in its
magnitude to the classical force
\begin{equation}
\label{quantumf}
\frac{\hbar^{2}}{m}c_{1}c_{2}=V_{1}[x(0)]=m\omega^{2}x(0).
\end{equation}
The LHS of this equation is the quantum force. Not only is the
quantum force not negligible, we see from the RHS of
eq.(\ref{quantumf}) that it depends linearly on the position. The
appearance of a significant quantum force in this most ``classical"
example can be viewed as a result of working with real trajectories.
In the BOMCA formulation with its complex trajectories, the quantum
force vanishes in this case, which is one of the strongest
motivations for studying BOMCA\cite{goldfarb}.

\section{Interference with real trajectories}
\label{inter}

\noindent As mentioned in the Introduction, the nodal problem is
currently the main obstacle to performing numerical calculations
using Bohmian mechanics. In this section, we present preliminary
results showing that an oscillatory wavefunction in close agreement
with the quantum result can be obtained using eqs.(\ref{N2}) with
real trajectories. Consider an initial Gaussian wavepacket
\begin{equation}
\label{initialwf}
 \psi(x,0)=\exp\left[-\alpha_{0}
 (x-x_{c})^{2}+\frac{i}{\hbar}p_{c}(x-x_{c})+\frac{i}{\hbar}\gamma_{0}\right],
\end{equation}
propagating in a Morse potential
\begin{equation}
\label{morse}
V(x)=A\left\{[1-\exp(-\beta x)]^{2}-1\right\}.
\end{equation}
The parameters of the initial Gaussian are $\alpha_{0}=0.5$,
$x_{c}=9.342$, $p_{c}=0$, $\gamma_{0}=-\frac{i
\hbar}{4}\ln(\frac{2\alpha_{0}}{\pi})$ where we take $m=\hbar=1$
(all quantities are given in atomic units). As for the Morse
parameters, $A=10.25$ and $\beta=0.2209$. The final propagation
time, $t_{f}=5.93$, is chosen so as to produce a strongly
oscillating pattern. Figure \ref{wf} depicts the potential and
wavefunctions.
\begin{figure}[t]
\begin{center}
\epsfxsize=9 cm \epsfbox{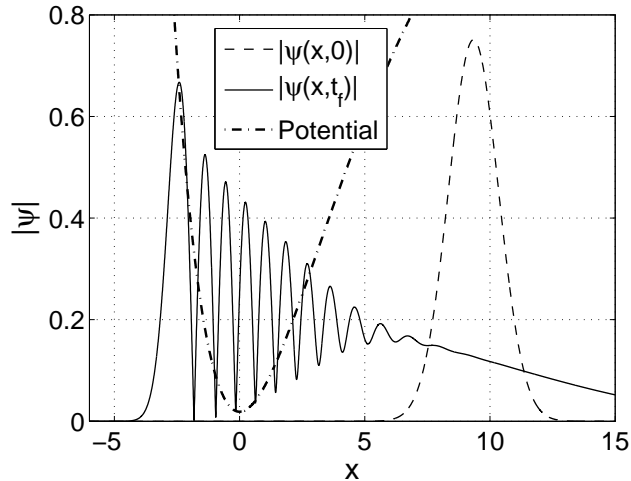} 
\end{center}
\caption{\label{wf} An initial Gaussian wavepacket propagating in a
Morse potential. The parameters of the system are given in the text.
The final propagation time $t_{f}$ equals roughly half of the
oscillation period of a classical particle located at the center of
the initial Gaussian, $x_{c}$. The exact wavefunction was calculated
using the split operator method with a Fourier basis.}
\end{figure}

In both DPM and BOMCA the velocity is taken to be a function of
$S_{1}$: in DPM, $v=\Re\left(\frac{S_{1}}{m}\right)$ while in BOMCA,
$v=\frac{S_{1}}{m}$. We now present an alternative procedure; we
take $v$ to be completely independent of $S_{1}$ and set a classical
equation of motion for $v$:
\begin{equation}
\label{v_revca} \frac{dv}{dt}=-\frac{V_{1}[x(t)]}{m}.
\end{equation}
Equation (\ref{v_revca}) is then supplemented to the set of
eqs.(\ref{N2}). Furthermore, we extend the freedom in the choice of
the velocity field to include the initial conditions for the
velocity, which we take to be
\begin{equation}
\label{initialv} v[x(0),0]=\frac{\Im\{
S_{1}[x(0),0]\}}{m}=\frac{2\alpha_{0}}{m}[x(0)-x_{c}];
\end{equation}
$S_{1}[x(0),0]$ is obtained by inserting eq.(\ref{initialwf}) into
eq.(\ref{initial}) and setting $n=1$. Equation (\ref{initialv})
defines real initial conditions; taken together with
eqs.(\ref{v_revca}) and (\ref{N2}a) this yields real classical
trajectories (note that if we had taken the initial velocity to be
complex, $v[x(0),0]=\frac{ S_{1}[x(0),0]}{m}$, we would have
obtained the BOMCA equations, eqs.(\ref{bomca_eqs})). The next step
is to solve eqs.(\ref{N2}) and (\ref{v_revca}) with initial
conditions (\ref{initial}) and (\ref{initialv}), respectively, for a
series of initial positions $\{x(0)\}$ in the vicinity of $x_{c}$.
The wavefunction at time $t_{f}$ at final position $x_{f}$ is given
by eq.(\ref{finalwf}), where $S_{0}[x(t_{f}),t_{f}]$ is the solution
of eq.(\ref{N2}b).

In fig.\ref{wyatt_2} we plot the trajectories obtained. The
trajectories can be divided into two overlapping groups that we
refer to as \textit{branches}. The first (reflected) branch (marked
as solid lines) is the locus of trajectories that have reached their
classical turning point and were reflected by the exponential
barrier of the potential. The second branch (marked as dashed lines)
is the locus of trajectories that by $t_{f}$ did not reach their
classical turning point.
\begin{figure}[t]
\begin{center}
\epsfxsize=9 cm \epsfbox{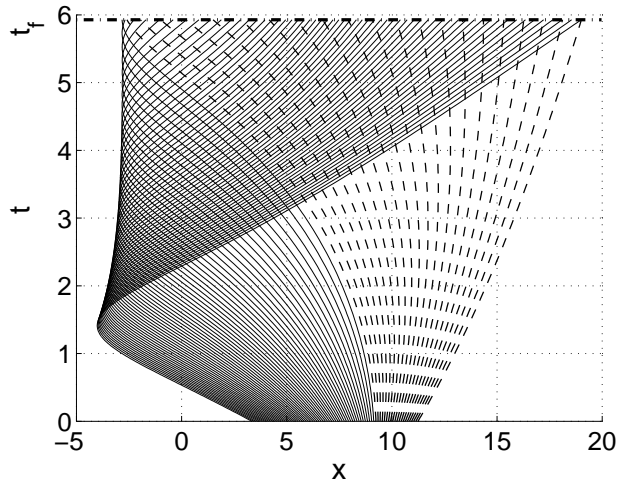}
\end{center}
\caption{\label{wyatt_2} Real classical trajectories obtained by
solving Newton's second law of motion (eq.(\ref{v_revca})) for a
Morse potential (eq.(\ref{morse})) with initial conditions
(\ref{initialv}). The parameters of the propagation are given in the
text. The final propagation time $t_{f}$ is marked explicitly. The
trajectories can be divided into two overlapping \textit{branches}.
The first branch (marked as solid lines) is the locus of
trajectories that have reached their classical turning point and
were reflected by the exponential barrier of the potential. The
second branch (marked as dashed lines) is the locus of trajectories
that by $t_{f}$ did not reach their classical turning point.}
\end{figure}
Thus, to an arbitrary final position $x$ ($x\gtrsim-2.8$) these
correspond two initial positions and two associated trajectories.

The question arises, if two trajectories and therefore two values of
$S_{0}(x,t_{f})$ correspond to each final position $x$, how should
we determine the wavefunction $\psi(x,t_{f})$? The clear distinction
between the two branches allows us to associate a wavefunction with
each branch, which we will call $\psi_{1}(x,t_{f})$ and
$\psi_{2}(x,t_{f})$. We apply the superposition principle to
reconstruct the wavefunction at $x$, obtaining
\begin{equation}
\psi(x,t_{f})=\psi_{1}(x,t_{f})+\psi_{2}(x,t_{f}).
\end{equation}
In fig.\ref{wyatt_1} we compare the exact wavefunction along with
$|\psi_{1}(x,t_{f})|$, $|\psi_{2}(x,t_{f})|$ and
$|\psi(x,t_{f})|=|\psi_{1}(x,t_{f})+\psi_{2}(x,t_{f})|$. For values
of $x$ larger than $\approx-2.8$ (the position of the maximum of
$|\psi(x,t_{f})|$), the superposition yields a surprisingly accurate
approximation of the oscillating wavefunction: even though the
wavefunctions $\psi_{1}(x,t_{f})$ and $\psi_{2}(x,t_{f})$ exhibit no
oscillations whatsoever, their superposition yields strong
oscillations and a \textit{node} near the maximum. Note that
$\psi_{2}(x,t_{f})$ provides the ``main" contribution to the final
wavefunction; $\psi_{1}(x,t_{f})$, which originates from the
reflected branch, contributes to $\psi(x,t_{f})$ only where the
wavefunction oscillates. As the classical turning point is
approached, the amplitudes arising from the different branches
diverge and the superposition of contributions appears to have no
physical significance.
\begin{figure}
\begin{center}
\epsfxsize=9 cm \epsfbox{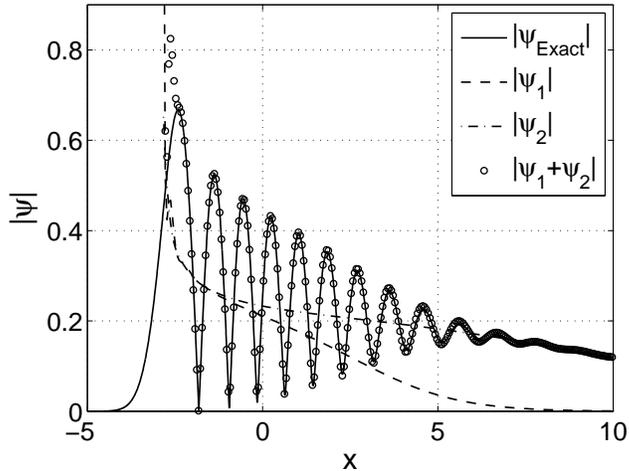} 
\end{center}
\caption{\label{wyatt_1} A comparison between the exact wavefunction
with the contribution of the first branch $|\psi_{1}(x,t_{f})|$, the
second branch $|\psi_{2}(x,t_{f})|$ and a superposition of both
$|\psi(x,t_{f})|=|\psi_{1}(x,t_{f})+\psi_{2}(x,t_{f})|$. For
$x\gtrsim-2.8$ (the position of the maximum of the exact
wavefunction) the superposition procedure yields a surprisingly
accurate approximation of the oscillating wavefunction: even though
the wavefunctions $|\psi_{1}(x,t_{f})|$ and $|\psi_{2}(x,t_{f})|$
exhibit no oscillations whatsoever. The procedure is undefined in
the classically forbidden region which the classical trajectories
cannot access. The exact wavefunction was calculated using the split
operator method with a Fourier basis.}
\end{figure}

As noted above, the BOMCA equations for $N=1$ are identical to those
of Generalized Gaussian Wavepacket Dynamics (GGWPD). In both cases,
the trajectories that are propagated obey classical equations of
motion but are complex. In this section we described a modification
of BOMCA in which the classical trajectories are taken to be real.
It is interesting to speculate if there might be a connection with
{\it thawed} Gaussian propagation, in which the equations of motion
are the same as those for GGWPD, but the trajectories are real.
Consistent with this conjecture is the observation that in thawed
Gaussian propagation there is no need for a root search, as is true
in the real-trajectory version of BOMCA. However, the correspondence
cannot be exact.  In the real-trajectory version of BOMCA, in
principle every point in coordinate space is propagated, whereas in
thawed Gaussian propagation the initial wavefunction is decomposed
into Gaussians and any decomposition that satisfies completeness is
allowed. Moreover, in real-trajectory BOMCA the trajectories have no
width and no functional form whereas in thawed Gaussian propagation
there is always some residual signature of the Gaussian
decomposition. For example, Gaussians that reach the turning point
have part of their amplitude extending into the classically
forbidden region, whereas in real-trajectory BOMCA such penetration
into the classically forbidden region is absent.

\section{Summary}
\label{summary}

\noindent We have presented a unified and compact derivation of
Bohmian methods that serves as a common starting point for several
approximations. In particular, the new approach was used to derive
and compare the DPM, BOMCA and ZEVCA methods. We focused on the role
of the quantum force in the DPM and BOMCA methods, and showed that
for $N=2$, the lowest order of truncation for which a quantum
potential is present, the BOMCA formulation is closer to the
classical equations of motion than DPM, although the trajectories
are complex. We also demonstrated that an interference pattern can
be obtained by superposing the contributions from real classical
trajectories in what is otherwise essentially a Bohmian formulation.

Since nodes in quantum mechanics arise from interfering amplitudes,
it is only natural to attempt to solve the nodal problem in Bohmian
mechanics by applying the superposition principle. This suggests
decomposing the wavefunction into two nodeless parts and propagating
each part separately using trajectories \cite{poirier,babyuk}.
However, since nodeless wavefunctions do not generally remain
nodeless, this approach generally requires a series of
time-dependent decompositions of the total wavefunction. Such
decompositions are numerically inconvenient, largely arbitrary, and
typically valid only for relatively short times. In the scheme we
presented, an oscillatory wavefunction is successfully decomposed
for all times considered into two nodeless, non-oscillatory parts,
$\psi_{1}$ and $\psi_{2}$. As such, the scheme represents a
promising new avenue for dealing with the challenging problem of
nodes in Bohmian mechanics. We are currently exploring an analogous
idea of applying the superposition principle to \textit{complex}
crossing trajectories in BOMCA\cite{goldfarb3} and in time-dependent
WKB\cite{goldfarb4}; the results are very promising and will be
published elsewhere. In addition, although we do not yet have final
results, it seems that path integral techniques can provide a
rigorous justification of the superposition principle in Bohmian
methods and the limits of its applications.

This work was supported by the Israel Science Foundation $(576/04)$.

%

\end{document}